\documentclass[12pt]{article}

\usepackage{bbm}

\newcommand{\be}{\begin{equation}}
\newcommand{\ee}{\end{equation}}
\newcommand{\ba}{\begin{eqnarray}}
\newcommand{\ea}{\end{eqnarray}}

\begin{document}

\title{\normalsize \hfill UWThPh-2003-45 \\[1cm] \LARGE
Quark mixing from softly broken symmetries}
\author{Walter Grimus\thanks{E-mail: walter.grimus@univie.ac.at} \\
\setcounter{footnote}{6}
\small Institut f\"ur Theoretische Physik, Universit\"at Wien \\
\small Boltzmanngasse 5, A--1090 Wien, Austria \\*[3.6mm]
Lu\'{\i}s Lavoura\thanks{E-mail: balio@cfif.ist.utl.pt} \\
\small Universidade T\'ecnica de Lisboa \\
\small Centro de F\'\i sica das Interac\c c\~oes Fundamentais \\
\small Instituto Superior T\'ecnico, P--1049-001 Lisboa, Portugal \\*[4.6mm] }

\date{23 December 2003}

\maketitle

\begin{abstract}
Quark flavor mixing may originate
in the soft breaking of horizontal symmetries.
Those symmetries,
which in the simplest case are three family $U(1)$ groups,
are obeyed only by the dimension-4 Yukawa couplings and lead,
when unbroken,
to the absence of mixing.
Their breaking may arise from the
dimension-3 mass terms of $SU(2)$-singlet vector-like quarks.
Those gauge-singlet mass terms break the horizontal symmetries
at a scale \emph{much higher} than the Fermi scale,
yet \emph{softly},
leading to quark mixing while the quark masses remain unsuppressed.
\end{abstract}

\newpage

\paragraph{Introduction} The seesaw mechanism \cite{seesaw}
is a very popular mechanism for explaining
the smallness of the neutrino masses.
Extending the lepton sector of the Standard Model (SM)
by right-handed neutrino singlets $\nu_R$,
one obtains the full Majorana mass matrix 
\be
\left( \begin{array}{cc} M_L & M_D^T \\
M_D & M_R \end{array} \right)
\label{D+M}
\ee
for the neutral components of the left-handed SM lepton doublets,
$\nu_L$,
together with the charge-conjugates of the neutrino singlets,
$\left( \nu_R \right)^c$.
If the scale inherent in $M_R$
is much larger than the scales of $M_L$ and $M_D$,
then eq.~(\ref{D+M}) leads to an effective mass matrix
\be
M_L - M_D^T M_R^{-1} M_D
\label{numass}
\ee
for the light neutrinos.

On the one hand this mechanism provides a rationale
for the smallness of the neutrino masses,
yet on the other hand it complicates the situation
with respect to lepton mixing.
This has now four sources: 
the charged-lepton mass matrix,
$M_L$,
$M_D$,
and $M_R$.
In \cite{GL01,oneloop} a scenario has been proposed
where the sole source of lepton mixing is $M_R$.
That scenario is based on the following assumptions: 
\begin{enumerate}
\renewcommand{\labelenumi}{\roman{enumi}.}
\item The lepton-family numbers $L_\alpha$
($\alpha = e,\mu,\tau$)
are conserved by all terms in the Lagrangian of dimension 4;
\item The $L_\alpha$ are broken \emph{softly}
by the $\nu_R$ Majorana mass terms in $M_R$,
which have dimension 3.
\end{enumerate}
Such a framework \cite{glashow} leads to a renormalizable theory,
and it has been shown \cite{oneloop} to be compatible
with all experimental constraints.
Within the above scenario several models
for explaining maximal atmospheric neutrino mixing
have been devised \cite{GL01,GL03,GL-$CP$,GLreview}.

In this letter we transfer the idea of obtaining lepton mixing
from softly broken lepton-family symmetries to the quark sector.
We add vector-like isosinglet quarks to the SM
and suggest that it is the dimension-3 mass terms of those quarks
which are responsible for mixing.

\paragraph{Extension of the SM quark sector}
$SU(2)$-singlet (isosinglet) vector-like quarks may be added
either to the SM down-type-quark sector,
or to the up-type-quark sector,
or to both simultaneously.
For definiteness we shall first assume that
they are added only to the down-type-quark sector;
an extension of this sector through vector-like isosinglet quarks
is motivated by the Grand Unified Theories
based on the Lie algebra $E_6$ \cite{E6}.
We add $n$ left-handed and $n$ right-handed quark $SU(2)$ singlets,
each of them with electric charge $-1/3$,
to the SM \cite{aguila,silva}.
We then have 3 left-handed quark doublets
$q_L = \left( p_L,\, n_L \right)^T$,
$n$ left-handed quark singlets $N_L$,
and $3+n$ right-handed quark singlets $N_R$;
there are also 3 right-handed quark fields $p_R$
with electric charge $+2/3$ \cite{BLS}.
The gauge symmetry allows the terms 
\be
-\mathcal{L} = \cdots +
\bar q_L \left( i \sigma_2 \phi^\ast \right) \Delta p_R +
\bar q_L \phi\, \Gamma N_R + \bar N_L M N_R + \mbox{H.c.},
\label{lagrangian}
\ee
where $\Delta$ is a dimensionless $3 \times 3$ matrix,
$\Gamma$ is a dimensionless $3 \times (3+n)$ matrix,
and  $M$ is a dimension-1 $n \times (3+n)$ matrix. 
For simplicity we have considered only one Higgs doublet,
but this framework is independent of the number of Higgs doublets.
Writing $M_p = v^\ast \Delta$ and $m = v \Gamma$,
where $v$ is the vacuum expectation value of $\phi^0$,
one obtains the mass terms
\be
\bar p_L M_p p_R
+ \left( \bar n_L,\, \bar N_L \right) \mathcal{M}\, N_R,
\ee
where the $\left( 3+n \right) \times \left( 3+n \right)$
matrix $\mathcal{M}$ is given by 
\be
\mathcal{M} = \left( \begin{array}{c} m \\ M \end{array} \right).
\label{M6x6}
\ee
For the discussion of quark mixing we must consider
\be
\mathcal{M} \mathcal{M}^\dagger = \left( \begin{array}{cc}
m m^\dagger & m M^\dagger \\ M m^\dagger & M M^\dagger \end{array} \right).
\label{Hmatrix}
\ee
It is natural to assume that $m$ is of the order
of the electroweak (Fermi) scale
while $M$ is of a much larger scale.
Then the effective mass-squared matrix for the light down quarks is
\cite{GL}
\be
D_\mathrm{light} =
m m^\dagger - m M^\dagger \left( M M^\dagger \right)^{-1}
\! M m^\dagger.
\label{light}
\ee
Notice the similarity  of the transition
from eq.~(\ref{Hmatrix}) to eq.~(\ref{light})
to the well-known seesaw transition
from eq.~(\ref{D+M}) to eq.~(\ref{numass}).
However,
eq.~(\ref{light}) does \emph{not} involve a true seesaw mechanism;
in general,
no suppression of the magnitude of the quark masses is produced
by eq.~(\ref{light}).

\paragraph{Softly broken quark-family numbers}
Equation~(\ref{light}) can be used \cite{$CP$}
to generate \emph{unsuppressed} $CP$ violation at low energies
from the spontaneous breaking of that symmetry in a heavy sector.
In this note we point out that
quark \emph{mixing} can also be generated from eq.~(\ref{light}),
if some horizontal symmetry is \emph{softly} broken in the matrix $M$.

As a first example,
let us assume that $n=3$;
then there are six $N_R$,
which we divide in two sets:
three $N_{R1}$ and three $N_{R2}$.
We furthermore assume that there are three family numbers $F_\alpha$
($\alpha = u, c, t$)
such that the fields $q_{L \alpha}$,
$N_{L \alpha}$,
$p_{R \alpha}$,
$N_{R1 \alpha}$,
and $N_{R2 \alpha}$ all have $F_\alpha = 1$,
and $F_\beta = 0$ for $\beta \neq \alpha$.
We require family-number conservation in all terms of dimension 4
in the Lagrangian.
Then,
\be
m = \left( m_1,\, m_2 \right),
\ee
where $m_1$ and $m_2$ are diagonal $3 \times 3$ matrices.
On the other hand,
\be
M = \left( M_1,\, M_2 \right),
\ee
and we allow the $3 \times 3$ matrices 
$M_1$ and $M_2$ to be non-diagonal,
since they correspond to dimension-3 (soft) terms in the Lagrangian.
Equation~(\ref{light}) now reads
\be
D_\mathrm{light} =
m_1 m_1^\dagger + m_2 m_2^\dagger
- \left( m_1 M_1^\dagger + m_2 M_2^\dagger \right)
\left( M_1 M_1^\dagger + M_2 M_2^\dagger \right)^{-1} \!
\left( M_1 m_1^\dagger + M_2 m_2^\dagger \right),
\label{fullight}
\ee
and $D_\mathrm{light}$ will in general be non-diagonal.
On the other hand,
the Yukawa-coupling matrix $\Delta$ in eq.~(\ref{lagrangian})
is diagonal,
and so is the up-type-quark mass matrix:
\be
M_p = \mathrm{diag} \left( m_u e^{i \phi_u},\, m_c e^{i \phi_c},\,
m_t e^{i \phi_t} \right),
\label{Mp}
\ee
where $m_{u,c,t}$ are the up-type-quark masses;
the $\phi_{u,c,t}$ are arbitrary phases,
which should be removed through rephasings of the $p_R$.

Both $M_1$ and $M_2$ must be present for non-trivial mixing to be obtained.
For instance,
if $M_1 = 0$ then eq.~(\ref{fullight}) gives
\be
D_\mathrm{light} =
m_1 m_1^\dagger + m_2 m_2^\dagger
- m_2 M_2^\dagger \left( M_2 M_2^\dagger \right)^{-1} \! M_2 m_2^\dagger =
m_1 m_1^\dagger,
\ee
which is diagonal.

Let us consider the mass terms involving the fields $n_L$.
Since $m_1$ and $m_2$ are diagonal,
there are three separate expressions
(one for each $\alpha$)
\be
\bar n_{L\alpha} \left[ \left( m_1 \right)_{\alpha\alpha} N_{R1\alpha} +
\left( m_2 \right)_{\alpha\alpha} N_{R2\alpha} \right].
\ee
We may rotate each pair $\left( N_{R1\alpha},\, N_{R2\alpha} \right)^T$
by a unitary $2 \times 2$ matrix such that
$\left( m_2 \right)_{\alpha \alpha}$ becomes zero.
Then $m_2 = 0$ and eq.~(\ref{fullight}) reads 
\be
D_\mathrm{light} =
m_1 m_1^\dagger
- m_1 M_1^\dagger
\left( M_1 M_1^\dagger + M_2 M_2^\dagger \right)^{-1}
\! M_1 m_1^\dagger.
\label{simplified}
\ee
One may further perform unitary transformations
of the $N_L$ and $N_{R2}$ in such a way that $M_2$
becomes diagonal and real non-negative
(bi-diagonalization of $M_2$).

It is interesting to notice that
a genuine seesaw mechanism for Dirac fermions 
may be obtained from eq.~(\ref{simplified})
if $M_1$ happens to be of a much larger order of magnitude than $M_2$;
we stress,
though,
that the inequality $M_1 M_1^\dagger \gg M_2 M_2^\dagger$ 
must hold in the basis where $m_2 = 0$.
Then we obtain the mass terms in leading order as \cite{japanese}
\be
- \bar n_L \left( m_1 M_1^ {-1} M_2 \right) N_{R2}
+ \bar N_L M_1 N_{R1}.
\ee

The diagonal elements of $M_1$ and $M_2$
conserve the quark-family numbers.
It is natural to assume that the off-diagonal elements
of those matrices
are small compared to the diagonal ones.\footnote{This assumption
may be misleading,
though,
since the down-type-quark masses have different orders of magnitude,
and therefore the diagonal elements of $M_1$ and $M_2$
are also likely to be widely different.}
This might give a rationale for rather small quark mixing,
in contrast to large lepton mixing.
Let us then assume that
\begin{description}
\item $m_2 = 0$;
\item both $m_1$ and $M_2$ are diagonal and real;
\item $M_1 = \hat M_1 + A$,
where $\hat M_1$ is diagonal and real
while $A$ has zeros in the diagonal
and its off-diagonal elements are small
compared to the diagonal elements of both $\hat M_1$ and $M_2$.
\end{description}
Then,
to first order in $A$ one obtains
\be
D_\mathrm{light} =
\frac{m_1^2 M_2^2}{\hat M_1^2 + M_2^2}
- \frac{m_1 \hat M_1}{\hat M_1^2 + M_2^2}\, A\,
\frac{m_1 M_2^2}{\hat M_1^2 + M_2^2} -
\frac{m_1 M_2^2}{\hat M_1^2 + M_2^2}\, A^\dagger\,
\frac{m_1 \hat M_1}{\hat M_1^2 + M_2^2}.
\ee

\paragraph{Less than 3 isosinglet quarks} In order to simplify
the above scheme one may allow for $n < 3$.
There will still be three $U(1)_\alpha$ groups
($\alpha = u, c, t$),
but one (if $n=2$) or two (if $n=1$) of these groups
act trivially on the $N_{R2}$.
For instance,
if $n=1$ then $M_1$ is a row vector with three elements
and $M_2$ is just a number;
$m_2$ is a column vector with three
elements.\footnote{The position of its non-zero element is
determined by the transformation property of $N_{R2}$ under the 
$U(1)_\alpha$ groups.}
As discussed before,
with a change of basis we achieve $m_2 = 0$.
Then,
with the notation
$m_1 = \mathrm{diag} \left( a_1,\, a_2,\, a_3 \right)$, 
$S^2 = M_1 M_1^\dagger + M_2 M_2^\dagger$
(now this is simply a number),
and $\ell_j = \left. \left( M_1 \right)_j \right/ S$,
from eq.~(\ref{simplified}) one obtains  
\be
\left( D_\mathrm{light} \right)_{ij} = 
a_i a_j^* \left( \delta_{ij} - \ell_i^* \ell_j \right).
\label{4x4}
\ee
From this equation we see that 
\be
\left( D_\mathrm{light} \right)_{12}
\left( D_\mathrm{light} \right)_{23}
\left( D_\mathrm{light} \right)_{31} =
- \prod_{j = 1}^3 \left| a_j \right|^2 \left| \ell_j \right|^2 
\ee
is real and negative.
Since the up-type-quark mass matrix in eq.~(\ref{Mp})
is diagonal in this basis,
$D_\mathrm{light} = V M_d^2 V^\dagger$,
where $M_d^2$ is the diagonal matrix of the squared
down-type-quark masses and $V$ is the
Cabibbo--Kobayashi--Maskawa (CKM) matrix.
Thus,
\be
\sum_{i,j,k = d,s,b} m_i^2 m_j^2 m_k^2
V_{ui} V_{ci}^\ast V_{cj} V_{tj}^\ast V_{tk} V_{uk}^\ast
\label{quantity}
\ee
is real and negative in this model.
This means that there is no $CP$ violation in the CKM matrix,
and thus $V$ may be set real.
Moreover,
given the known hierarchy of the down-type-quark masses
and of the CKM matrix elements,
the quantity in eq.~(\ref{quantity}) is approximately equal to
\be
m_b^4 V_{tb}^2 V_{ub} V_{cb}
\left( m_b^2 V_{ub} V_{cb} + m_s^2 V_{us} V_{cs} \right),
\ee
which must be negative.
This implies that the CKM-matrix Wolfenstein parameter
$\rho$ \cite{wolfenstein}
is negative in this model.
A negative $\rho$
is experimentally disfavored in the SM---see for
instance \cite{BLS}---but
here things are more complicated,
since this model has flavor-changing quark currents
coupling to the $Z^0$ \cite{aguila,silva},
and, moreover, it must be endowed with some extra structure,
like for instance several Higgs doublets,
in order to account for $CP$ violation.

\paragraph{Non-Abelian symmetries}
Another possibility consists in assuming that 
the horizontal symmetry is non-Abelian,
instead of being the product of three $U(1)_\alpha$ groups.
This is interesting since $m_1$ and/or $m_2$
may then become more constrained,
in particular two or three of their diagonal elements may become equal.
Relevant discrete groups might for instance be the
dihedral groups \cite{Delta},
$S_4$,
or $A_4$,
which have three-dimensional irreducible representations (irreps),
or $S_3$,
or $D_4$,
which have two-dimensional irreps.

For instance,
the group of the permutations of four objects,
$S_4$,
has a three-dimensional faithful irrep $\underline{3}$ generated by 
\be
(13) \to 
\left( \begin{array}{ccc} 0 & 0 & -1 \\ 0 & 1 & 0 \\ -1 & 0 & 0
\end{array} \right), \quad
(24) \to 
\left( \begin{array}{ccc} 0 & 0 & 1 \\ 0 & 1 & 0 \\ 1 & 0 & 0
\end{array} \right), \quad
(12) \to 
\left( \begin{array}{ccc} 1 & 0 & 0 \\ 0 & 0 & -1 \\ 0 & -1 & 0
\end{array} \right),
\ee
where $(13)$,
$(24)$,
and $(12)$ are transpositions.
The product $\underline{3} \otimes \underline{3}$
contains only one invariant of $S_4$;
if $\psi_j$ ($j=1,2,3$) and $\psi_j^\prime$ ($j=1,2,3$)
are two triplets of $S_4$,
then $\psi_1 \psi_1^\prime + \psi_2 \psi_2^\prime + \psi_3 \psi_3^\prime$
transforms trivially under $S_4$.
One may then postulate that the three $q_{L\alpha}$,
the three $N_{R1\alpha}$,
and the three $N_{R2\alpha}$ are all triplets of $S_4$;
if the Higgs doublets are invariant under $S_4$,
this immediately makes $m_1$ and $m_2$ proportional to the unit matrix,
and thus diagonal,
even in the absence of any $U(1)_\alpha$ symmetries.
As for $M_1$ and $M_2$,
they are arbitrary since we allow $S_4$ to be softly broken.

Another example is the symmetry $D_4$,
which has a two-dimensional faithful irrep $\underline{2}$
given by the matrices
\be
\left( \begin{array}{cc} 1 & 0 \\ 0 & -1 \end{array} \right),
\quad
\left( \begin{array}{cc} 0 & 1 \\ 1 & 0 \end{array} \right),
\ee
and all the products thereof.
The product $\underline{2} \otimes \underline{2}$
contains only one invariant $\underline{1}$ of $D_4$;
if $\left( \psi_1,\, \psi_2 \right)$
and $\left( \psi_1^\prime,\, \psi_2^\prime \right)$
are two doublets of $D_4$,
then $\psi_1 \psi_1^\prime + \psi_2 \psi_2^\prime$ 
is invariant under $D_4$.
Thus,
if the three $q_{L \alpha}$ transform as
$\underline{2} \oplus \underline{1}$
and if $n=1$ with the four $N_R$ transforming as 
$\underline{2} \oplus \underline{1} \oplus \underline{1}$,
then, choosing a basis where $m_2 = 0$, 
$\mathcal{M}$ will be as described before 
in eq.~(\ref{4x4}) but with $a_1 = a_2$.

A disadvantage of non-Abelian symmetries,
though,
is that one must allow for isosinglet vector-like quarks
in both the down-type- and up-type-quark sectors.
Indeed,
since the $q_{L\alpha}$ transform as a 
multi-dimensional representation of some discrete
(or continuous)
non-Abelian group,
the absence of soft breaking of that group in one of the sectors
leads to degenerate quarks in that sector.
In order to avoid having two quarks with equal masses,
the discrete group
must be (either softly or spontaneously) broken
in both the up-type-quark and down-type-quark sectors.

\paragraph{A particular case} Suppose for instance
that $n=2$ and that there is a softly broken symmetry $S_4$ such that
\be
\mathcal{M} = \left( \begin{array}{cc} m & 0 \\ M & N \end{array} \right),
\ee
where $m$ is proportional to the unit matrix
and (without loss of generality) real,
while $M$ is an arbitrary $2 \times 3$ matrix
and $N$ an arbitrary $2 \times 2$ matrix.
Then,
\be
D_\mathrm{light} = m^2 \left[ \mathbbm{1} - 
M^\dagger \left( M M^\dagger + N N^\dagger \right)^{-1} \! M \right].
\label{D}
\ee
We assume that the two row vectors in $M$
are linearly independent.
Then there is a column vector
$v_3$---unique up to a phase---such that $M v_3 = 0$.
It is clear that $D_\mathrm{light} v_3 = m^2 v_3$.
There is a $2 \times 2$ unitary matrix $U$
and a $3 \times 3$ unitary matrix $W$ such that
\be
M = U \left( \begin{array}{ccc} \mu_1 & 0 & 0 \\
0 & \mu_2 & 0 \end{array} \right) W^\dagger,
\ee
with real and positive $\mu_{1,2}$.
Then,
with $\hat \mu = \mathrm{diag} \left( \mu_1,\, \mu_2 \right)$,
\be
M M^\dagger = U \hat \mu^2 U^\dagger. 
\ee
It is easy to see that
\be
M^\dagger \left( M M^\dagger + N N^\dagger \right)^{-1} \! M = 
W \left( \begin{array}{cc} 1 & 0 \\
0 & 1 \\ 0 & 0 \end{array} \right) 
\left( \mathbbm{1} + B \right)^{-1} 
\left( \begin{array}{ccc} 1 & 0 & 0 \\
0 & 1 & 0 \end{array} \right) W^\dagger,
\ee
where
\be
B = {\hat \mu}^{-1} U^\dagger N N^\dagger U {\hat \mu}^{-1}.
\ee
Let $x_1$ and $x_2$ be the eigenvectors
of the $2 \times 2$ Hermitian matrix $B$,
corresponding to the (real and positive)
eigenvalues $\lambda_1$ and $\lambda_2$,
respectively.
Write $W = \left( v_1,\, v_2,\, v_3 \right)$,
where $\{ v_1, v_2, v_3 \}$ is an orthonormal system of 3-vectors.
(Obviously,
$v_3$ is the vector satisfying $M v_3 = 0$.)
We define
\be
y_j = \left( x_j \right)_1 v_1 + \left( x_j \right)_2 v_2
\quad \mathrm{for}\ j = 1, 2.
\ee
Then,
\be
D_\mathrm{light} y_j = \frac{m^2 \lambda_j}{1+\lambda_j}\, y_j
\quad \mathrm{for}\ j = 1, 2.
\ee
We have thus found that $y_1$,
$y_2$,
and $v_3$ are the eigenvectors of $D_\mathrm{light}$
corresponding to the eigenvalues
$m^2 \lambda_1 \left/ \left( 1 + \lambda_1 \right) \right.$,
$m^2 \lambda_2 \left/ \left( 1 + \lambda_2 \right) \right.$,
and $m^2$,
respectively.

Since $N$ is arbitrary,
$N N^\dagger$ is an arbitrary Hermitian matrix,
and $B$ too can be taken as a general Hermitian positive
$2 \times 2$ matrix.
Therefore,
$\lambda_1$ and $\lambda_2$ are arbitrary.
Also,
$m^2$ is unconstrained by the symmetry $S_4$.
Finally,
since $M$ is arbitrary,
$W$ too is an arbitrary $3 \times 3$ unitary matrix,
which means that $y_1$,
$y_2$,
and $v_3$ are arbitrary.
We thus conclude that the matrix $D_\mathrm{light}$ in eq.~(\ref{D})
is a completely general $3 \times 3$ mass-squared matrix.
This means that our simple case with $n=2$
and softly broken $S_4$ symmetry
leaves $D_\mathrm{light}$ unconstrained.

\paragraph{Conclusions} In this letter we have suggested that
the Yukawa couplings of an extension of the SM
with isosinglet vector-like quarks may respect a horizontal symmetry
which would lead to trivial quark mixing.
However,
that horizontal symmetry is \emph{softly} broken
\emph{at a high scale} (much above the Fermi scale)
by the gauge-invariant mass terms of the vector-like quarks.
We have stressed that this leads to
\emph{unsuppressed} quark mixing at low energies,
even though that mixing arises from the heavy sector of the theory.
The number of vector-like quarks is largely arbitrary,
and the horizontal symmetry may either consist in the product
of three $U(1)$ quark-family groups
or it may be a discrete or continuous non-Abelian group
implying flavor-diagonal Yukawa couplings.

\paragraph{Acknowledgement} The work of L.L.\ was supported
by the Portuguese \textit{Funda\c c\~ao para a Ci\^encia e a Tecnologia}
under the contract CFIF-Plurianual.

\end{document}